\documentclass[12pt]{article}

\usepackage{graphicx}
\usepackage{graphics}
\usepackage[margin=1in]{geometry}
\usepackage{amssymb}
\usepackage{amsmath}
\usepackage[table]{xcolor}
\usepackage{setspace}
\usepackage[round]{natbib}
\usepackage{url}
\usepackage{authblk}

\newcommand{\ellhat}{{{\ell}_n}}

\newcommand{\what}{{w_n}}
\newcommand{\uhat}{{u_n}}

\newcommand{\thetahat}{\hat{\theta}}

\newcommand{\sumin}{\sum_{i=1}^n}

\newcommand{\sumjm}{\sum_{j=1}^{m}}

\newcommand{\arrowp}{\overset{p}{\rightarrow}}
\newcommand{\arrowd}{\overset{d}{\rightarrow}}
\newcommand{\thetaxi}{\theta_\xi^\ast}

\newcommand{\xsubi}{X^{(i)}}

\newtheorem{theorem}{Theorem}[section]

\newtheorem{proposition}[theorem]{Proposition}

\begin{document}
\doublespacing

% \markboth{D. Ferrari}{Reliable composite likelihood inference}

\title{Reliable inference for complex models by discriminative composite likelihood estimation}

\author[1]{Davide Ferrari\thanks{Address: Richard Berry Building, University of
Melbourne, Parkville, 3010, VIC, Australia; Phone: +61 383446411; E-mail: dferrari@unimelb.edu.au}}
\author[2]{Chao Zheng}
\affil[1,2]{School of Mathematics and Statistics\\ University of Melbourne}

\maketitle

\begin{abstract}
Composite likelihood estimation has an important role in the analysis of multivariate data for which the full likelihood function is intractable.
An important issue in composite likelihood inference is the choice of the weights associated with lower-dimensional data sub-sets, since
the presence of incompatible sub-models can deteriorate the accuracy of the resulting estimator. In this paper, we introduce a new approach for
simultaneous parameter estimation  by tilting, or re-weighting, each sub-likelihood component called discriminative composite likelihood estimation
(D-McLE). The data-adaptive weights maximize the composite likelihood function, subject to moving a given distance from uniform weights; then, the
resulting weights can be used to rank  lower-dimensional likelihoods in terms of their influence in the composite likelihood function. Our analytical
findings and numerical examples support the stability of the resulting estimator compared to estimators constructed using standard composition strategies based
on uniform weights. The properties of the new method are illustrated through simulated data and real spatial data on multivariate precipitation extremes.
\end{abstract}


\section{Introduction}

While  likelihood-based inference is central to modern statistics, for many multivariate problems the full likelihood function
is impossible to specify or its evaluation involves a
prohibitive computational cost. These limitations have motivated the development of
composite likelihood approaches, which avoid the full likelihood by
compounding a set of low-dimensional likelihoods into a surrogate criterion function. Composite likelihood inference
have proved useful in a number of fields, including geo-statistics, analysis of spatial
extremes, statistical genetics, and longitudinal data analysis. See \cite{Varin&al11}
for a comprehensive survey of composite likelihood theory and applications. \cite{larribe2011composite} review several applications in genetics.

Let $X$ be a $d \times 1$  random vector and
$f(x| \theta)$ be the assumed density model for $X$, indexed by the parameter $\theta \in
\Theta \subseteq \mathbb{R}^p$, $p
\geq 1$. Suppose that the full likelihood function, $L(\theta|x) \propto f(x| \theta)$, is
difficult to
specify or compute, but we can specify low-dimensional distributions with one, two, or
more
variables. Specifically, let $\{ Y_j,  j=1,\dots, m\}$ be a set of marginal or
conditional low-dimensional
variables constructed from $X$ with associated likelihoods $L_j(\theta|y_j) \propto
f_j(y_j|\theta)$, where  $f_{j}(\cdot|\theta)$, $\theta \in \Theta$ denotes the $j$th low-dimensional density model
for $Y_j$. The low-dimensional variables $\{Y_j \}$ are user-defined and
could be constructed by taking marginal models, like $X_1, \dots, X_d$, pairs like $(X_1,
X_2)$, or conditional variables like $(X_1, X_2)|X_2$. The overall structure of such lower-dimensional models is sometimes referred as to composite likelihood
design \citep{Lindsay&al11} and
its choice is often driven by computational convenience. For example, if $X$ follows a $d$-variate normal distribution $N_d(0, \Sigma)$, the full likelihood
is hard to compute when $d$ is large due to inversion of $\Sigma$, which involves  $O(d^3)$ operations. In contrast, using sub-models for variable
pairs $(X_k,
X_{k^\prime})$, $1 \leq k < k^\prime \leq d$, can reduce the computational burden since it involves simply inverting $2 \times 2$ partial
covariance matrices.

Following \cite{Lindsay88}, we define
the composite likelihood function  by
\begin{equation} \label{Lcl}
CL(\theta|w, x) = \prod_{j=1}^{m} f_j( y_j| \theta)^{w_j},
\end{equation}
where $\{w_j, j=1,\dots,m\}$ are non-negative weights, possibly depending on $\theta$. A well-known issue in composite likelihood estimation is the selection of
the weights, as their specification plays a crucial role in determining both efficiency and reliability of the resulting composite likelihood
estimator  \citep{Lindsay88, Joe&Lee09, Cox&Reid04, Varin&al11, Xu&Reid11}. Despite the importance of the weights, many statistical and computational
challenges still hinder their  selection \citep{Lindsay&al11}.

This paper is concerned with the aspect of stability of composite likelihood selection. Stability occurs when the maximizer of the overall composite
likelihood function
$L(\theta|w)$  is not overly affected by the existence of locally optimal parameters that work only for a relatively
small portion of such sub-sets, say $Y_1, \dots, Y_{m^{\ast}}$, $m^\ast<m/2$. The presence of such local optima
arises from the incompatibility between the assumed full-likelihood model and the $m^\ast$ lower dimensional models.  For example, suppose that the true
distribution of $X$ is  a
$d$-variate normal distribution with zero mean vector, unit variance and correlations $2\rho_0$ for all variable pairs, while the true correlation is $\rho_0$
for some small fraction of the $d(d-1)/2$ pairs. If one mistakenly assumes that all correlations are equal to $\rho_0$, both maximum likelihood and pair-wise
likelihood estimators with uniform weight, $w_j = 1/m$, $j=1,\dots,m$, are not consistent for
$\rho_0$ in this situation. Other examples of incompatible models are given in \cite{Xu&Reid11}. In applications, model compatibility is  hard to detect,
especially when $m$ is large, so incompatible sub-models are often included in the composite likelihood function with detrimental effects on the accuracy of the
global composite likelihood estimator.

Motivated by the above issues, we introduce the discriminative maximum composite likelihood estimator (D-McLE), a new methodology for reliable
likelihood composition and simultaneous parameter estimation. The new approach
computes smooth weights by maximizing the
composite likelihood function for a sample of observations subject to moving a given distance,
say $\xi$, from uniform weights. The D-McLE is regarded as a generalization of the traditional McLE. If $\xi=0$ the D-McLE is exactly the
common composite likelihood estimator with
uniform weights. When $\xi>0$, incompatible sub-models are down-weighted, thus resulting in
estimators for $\theta$ with bounded worst-case bias. Our analytical findings and simulations support the validity of the proposed method compared
to classic composite likelihood estimators with uniform weights. The new framework is illustrated through estimation of
max-stable models, which have proved useful for describing extreme environmental occurrences as hurricanes, floods and storms \citep{davison2012}.

The proposed procure would be useful in two respects. First, the resulting weights would be a valuable diagnostic tool for composite likelihood selection.
Small weights would signal  suspicious models, which could be further examined leading to improved assumptions. Conversely, the method can be employed to
identify influential data sub-sets for many types of composite likelihood estimators. Second, the  estimates obtained by such method would be trustworthy at
least for the bulk of the data sub-sets models (which are compatible with model assumptions). Clearly, assigning the same weight to all the models including the
ones in strong disagreement with the majority of data would lead to biased global estimates, which can be an untrustworthy representations of the entire
data-set.

The proposed method is a type of data tilting, a general technique which involves replacing uniform weights with more general weights. To
our knowledge, this is the first work that introduces tilting for lower-dimensional data sub-sets within the composite likelihood framework. In robust
statistics, tilting has been typically employed to robustify parametric estimating equations, or to obtain natural data  order in terms of their influence
\cite{Choi00}. Tilting has also been used to obtain measures of outlyingness and influence of data-subsets; e.g., see \cite{hall99, critchley04, lazar05, camponovo12}.  \cite{genton2014} use a tilting approach in the context of multivariate functional
data to ranking influence of data subsets.

The rest  the paper is organized as follows. In Section \ref{sec2}, we describe the new methodology for simultaneous likelihood
selection/estimation; we give an efficient algorithm and introduce
the compatibility plot, a new graphical tool to assess the adequacy of the sub-models. In Section \ref{sec3},  we study the
properties of the new estimator and give its limit distribution.  In Section \ref{examples}, we provide simulated examples in finite samples confirming our
theoretical findings. In Section \ref{sec5}, we illustrate the new procedure to the Tasmanian rainfall spatial data  on multivariate precipitation extremes.
In Section \ref{sec:conclusion}, we conclude and discuss possible extensions for $m \rightarrow \infty$. Proofs of technical results are deferred to a
separate appendix.

\section{Methodology} \label{sec2}

\subsection{Composite likelihood selection} \label{sec:2.1}

 Given independent observations $X^{(1)},\dots, X^{(n)}$ from the true distribution
$G(x)$, we construct the set of marginal or conditional low-dimensional
observations $Y^{(1)}_j, \dots, Y_j^{(n)}$, $j=1, \dots, m$, and
define the weighted composite log-likelihood function
\begin{equation}  \label{comp_lik}
\ellhat(\theta| w) \equiv
\sum^m_{j=1} w_{j} \ellhat_{j}(\theta)
\equiv  \sum^m_{j=1} \dfrac{w_j}{n} \sumin
\log f_{j}(Y_j^{(i)}| \theta),
\end{equation}
where $w=(w_1, \dots, w_{m})^T \in [0,1]^{m}$ are constants playing the role of
importance weights. The weight $w_j$  characterizes the impact of the $j$th
sub-likelihood,
$$
\ellhat_{j}(\theta) \equiv n^{-1} \sumin \log f_j(y_j|\theta),$$
on the overall composite likelihood function $\ellhat(\theta| w)$. We define incompatibility  by assuming there is a global parameter, say $\theta_0 \in
\Theta$, which suits  most sub-models.
Specifically, we assume partial models $Y_j \sim f_j(y_j|\theta_j)$, where $\theta_j \neq \theta_0$ if $j \leq m^\ast < m/2$
(incompatible models) and  $\theta_{j} = \theta_0$, if   $m^\ast < j \leq m$ (compatible models).

% In addition,  compatible sub-models are regarded as
% superior in terms of their Fisher information. Thus, we assume $-E \nabla \ellhat_{j}(\theta_j) \le -E \nabla \ellhat_{k}(\theta_k)$, where the
% inequality is understood  in the Loewner partial order sense and it is satisfied for all $j\leq m^\ast$ and $m^\ast < k \leq m$.

Next, we introduce the D-McLE procedure for simultaneous discrimination of discordant models and parameter estimation. We propose to
select the weight $w_j$ to be small when, for a value of $\theta$ that is appropriate for the majority
of the data
sub-sets, the sub-likelihood function for the $j$th data sub-set,  $\ellhat_{j}(\theta)$, is
small. To this end, $w$ is regarded as a discrete distribution on
$m$ points and the discrepancy between $w$ and the uniform distribution
$w_{unif}=(1/m, \dots, 1/m)$ is measured by the Kullback-Leibler divergence
\begin{equation} \label{KL}
D_{KL}(w, w_{unif}) =  \sumjm w_j \log(m w_j),
\end{equation}
where $0 \leq D_{KL}(w, w_{unif})  \leq \log m$. For a given parameter $\theta$, data-dependent weights $w_n = w_n(\theta)$
are then chosen by solving the following program
\begin{equation}\label{estimator}
            \max_{w} \left\{ \ellhat(\theta| w) \right\}, \
\
\text{s.t.: } \ \  D_{KL}(w, w_{unif}) = \xi, \ \sumjm w_j=1.
\end{equation}
Finally, the D-McLE, denoted by $\theta = \thetahat_\xi$, is then defined as the maximizer of the composite log-likelihood function
$$
 \ellhat(\theta)  \equiv \ellhat(\theta| w_n(\theta))
$$
where $w_n(\theta)=(w_{n1}(\theta), \dots, w_{nm}(\theta))^T$ is the vector of data-dependent weights. Equivalently, $\thetahat_\xi$ can be obtained by
computing the profiled estimator $\hat{\theta}(w)$ by maximizing  $\ell_n(\theta|w)$ for a given weight and then solve (\ref{estimator}) with $\theta =
\hat{\theta}(w)$.

The composite likelihood estimator $\thetahat_\xi$ entails moving away
from uniform weights in the direction that emphasizes the contribution of the most useful
data sub-sets.  If $\xi>0$, the relative importance of
the sub-likelihoods that are incompatible with the data is diminished in the composite likelihood equation
(\ref{comp_lik}).
The special case when $\xi=0$ corresponds to the composite likelihood estimator
with uniform weights $w=w_{unif}$. Thus, all the data sub-sets are regarded as equally compatible. Other divergence measures may be considered in place
of the Kullback-Leibler divergence (\ref{KL}), which could be useful in particular estimation setups, although these are not pursued
in this paper. The Kullback-Leibler divergence, however, has the advantage that allows one or more zero weights, and gives automatically nonnegative wights
without imposing additional constraints by some algorithm to ensure this property. For example, when $m$ is  very large it could be useful to modify
$D_{KL}(w)$ to promote sparsity, i.e. select relatively a large number weights that are exactly zero.

\subsection{Data-adaptive weights and parameter estimation}
\label{computing1}

The program in (\ref{estimator}) is solved by
maximizing the Lagrangian function
\begin{equation} \label{lagrangian}
h(w, \lambda_1, \lambda_2| \theta) = \sumjm w_{j} \ellhat_{j}(\theta)
 +
\lambda_1 \left\{ D_{KL}(w, w_{unif}) - \xi \right\}  + \lambda_2
\left(
\sumjm  w_j - 1\right),
\end{equation}
where $\lambda_1$  and $\lambda_2$ are Lagrange multipliers. It is easy to see that the solution to (\ref{lagrangian}) has the form
\begin{equation}\label{weights}
w_{nj}(\theta) \equiv  \alpha_2
\exp\{\alpha_1 \ellhat_j(\theta) \}, \ \ j=1,\dots,m,
\end{equation}
where $\alpha_1$ and $\alpha_2$ depend on the Lagrange multipliers $\lambda_1$ and $\lambda_2$. From the two
constraints in (\ref{estimator}),
$\alpha_1\equiv \alpha_1(\theta)$ and   $\alpha_2\equiv
\alpha_2(\theta)$ are
obtained by
solving
\begin{equation}\label{eq3}
\xi = \alpha_1 \dfrac{\sumjm \exp\{\alpha_1
\ellhat_j(\theta) \} \ellhat_j(\theta)
}{\sumjm \exp\{\alpha_1 \ellhat_j(\theta) \} } - \log \sumjm  \exp\{\alpha_1
\ellhat_j(\theta)
\}   + \log m,
\end{equation}
and $\alpha_2 = 1/ \sumjm \exp\{\alpha_1 \ellhat_j(\theta)\}$.  The D-McLE  $\thetahat_\xi$ is
then computed by maximizing $\ellhat(\theta) \equiv \ellhat(\theta| \what(\theta))$.

Lemma 1 in the appendix shows that computing the D-McLE, $\hat{\theta}_\xi$, is equivalent to solving the estimating equations
\begin{align} \label{score}
\uhat(\theta) \equiv \nabla_\theta \ellhat(\theta) = \sumjm \what_j(\theta)
\uhat_j(\theta) = 0,
\end{align}
where $\uhat_j(\theta) \equiv n^{-1} \sumin u_j(Y_j^{(i)},\theta)$ denotes the partial score function corresponding to the $j$th data subset. Thus,
$\uhat(\theta)$ is a weighted estimating equation involving
the partial scores with weights depending on the data and $\theta$.  A small weight $w_{nj}$ implies a modest contribution of the $j$th score, $u_{nj}$, to the
overall composite likelihood equation. The  constant $\xi$ is regarded as a  stability parameter which can be used to control for the relative
impact of the incompatible lower-dimensional likelihoods.  Particularly, if $\xi$ is large incompatible models will receive a low
weight, with a relatively small effect on the final parameter estimates. If $\xi=0$, all the sub-models are treated equally in terms of the impact
of corresponding sub-likelihoods in $\uhat(\theta)$.

Equation (\ref{score}) highlights the  resemblance to estimating functions of classic robust M-estimators, whose main aim is to
reduce the influence of outliers in the full likelihood function. Indeed, the approach followed here coincides with the robust estimation approach by
\cite{Choi00} in the  particular case where: $n=1$, $Y_1, \dots, Y_m$ are independent  and all sub-models $f_j$, $j=1,\dots, m$ are all identical to the full
likelihood  model, $f$.
In general, however, the D-McLE  is  very different from \cite{Choi00} and other similar robust methods. The main difference is that the weights $\{w_{nj}\}$
in (\ref{weights})  refer to variables $Y_1, \dots, Y_m$, which are constructed by taking sub-sets of the original vector $X$ and are possibly correlated; in
robust M-estimation  weights refer to independent observations on the original vector $X$. Thus, in our approach $n$ observations
corresponding to the  $j$th data sub-set, namely $Y^{(i)}_j$, $i=1,\dots,n$, receive the same weight, $w_{jn}$.  This reflects our need to control for the
incompatibility of a portion of the sub-models, say $f_1, \dots, f_{m^\ast}$, $m^\ast<m$, rather than reducing the effect of outlying observations with respect
to the full  model $f$.

\subsection{Computing} \label{computing}

The form of equation (\ref{score}) suggests a simple  algorithm to
simultaneously compute weights and parameter estimates. At each step of the algorithm, we update  weights based on previous parameter estimates and then
compute  a fresh parameter estimate using the new weights. Starting from an initial
estimate, $\thetahat^{(0)}$, we compute:
\begin{align} \label{alg}
\thetahat^{(t)} = \left\{ \theta: \  0 = \sumjm \hat{w}_{j}(\thetahat^{(t-1)})
 \uhat_j( \theta) \right\},  \ \ t\geq 1,
\end{align}
until convergence is reached. We consider a relative convergence
criterion on the weights and stop iterating when $\Vert \what_{j}^{(t+1)}- \what_{j}^{(t)}\Vert/ \Vert \what_{j}^{(t)} \Vert <
\varepsilon$, where $\varepsilon>0$ is some tolerance level.  A 
practical advantageis that (\ref{alg}) is easy to implement when a basic composite likelihood estimator with fixed weights is already 
available. 

In our numerical studies, the algorithm gave satisfactory performances. In all our examples
convergence was reached in a few iterations and we noted that the computational cost does not increase
much as $m$ grows. This behavior  makes the proposed algorithm well-suited to high-dimensional problems with a large number of sub-likelihoods and is shared by 
analogous iteratively re-weighted algorithms for M-estimation with well-established theory (e.g. see \cite{arslan2004convergence}).   Although we do not 
offer theoretical insight on the general theoretical  behaviour of our algorithm, convergence results may be derived following an argument analogous to 
\cite{basu2004iteratively}  in the context of iteratively reweighted procedures for minimum divergence estimators.

\subsection{Compatibility profile plots (CPPs)} \label{cpp}

Let  $\Pi(\xi) = (p_1, \dots, p_m)$ be the arrangement of indices $\{1,\dots, m\}$ implied
by  $\what_{p_1}(\thetahat_\xi) < \dots <
\what_{p_m}(\thetahat_\xi)$, where $\what_{j}(\thetahat_\xi)$, $j=1,\dots,n$, are data-dependent weights
computed by the algorithm in Section \ref{computing}.
The ordering $\Pi(\xi)$ induces an
importance ranking for the sub-models in
terms of  their compatibility with the true distribution generating the data.
Based on this ranking, a graphical tool is introduced, called a compatibility
profile plot (CPP). The CPP traces the
fitted weights, $\what_{j}(\thetahat_\xi)$, $j=1,\dots, m$, as $\xi$ moves away from zero and can be
used to inspect the compatibility of individual sub-likelihoods. For instance, a sharp
decrease of the first $m^\ast$ weights from uniform weights $w_{unif}=(1/m, \dots, 1/m)$, suggests that the first $m^\ast$ sub-likelihoods are likely to be
misspecified and a
different model should be used for such components.
The weights  often exhibit diverging trajectories (see for example  Figure \ref{fig1}) which may be used to determine a suitable value for the parameter $\xi$.
For example, the plots help us pick a value of $\xi$ corresponding to a sufficient degree of separation between compatible and incompatible models. Eventually,
$\xi$ reaches an equilibrium point where the trajectories are maximally separated. After equilibrium, $m-1$ weights cluster together again as they tend to 0,
where a single weight converges to 1.

\subsection{Selection of  $\xi$}

The stability parameter $\xi$ tunes the extent to which we down-weight incompatible models, which is important to discuss. One approach is to select the tuning
constant $\xi$ closest
to 0 (i.e., closest to uniform weights) such that the point estimates of the parameters of interest are
sufficiently stable. If all the sub-likelihoods are compatible, $\xi=0$ already gives stable estimates and moving away $\xi$ is expected to have little
impact on the estimates. In the presence of incompatible sub-likelihoods, values of $\xi$ close to $0$ tend give unstable estimates in terms of bias and
variance, so we move $\xi$ away from $0$ until  stability is reached. For example in Figure \ref{fig1} (right), the correlation
estimator  $\hat{\rho}_\xi$ is far from the true correlation value of 0.5 when $\xi=0$. As $\xi$ moves away from zero, $\hat{\rho}_\xi$ changes rapidly until
 stability is reached when $\xi=0.51$. The above discussion suggests a simple data-driven procedure to select $\xi$:
\begin{enumerate}
\item[(1)] Define an equally spaced grid $0 = \xi_{0} < \xi_1 < \xi_2 < \dots < \xi_r \leq \log m$.
\item[(2)] Starting from $\xi_0$ compute the correspondent point estimates, $\thetahat_{\xi_i}$, $i=0,\dots,r$.
\item[(3)] Select the optimal value using the stopping rule $\hat{\xi} =\{ \min \ \xi_i: \Vert \thetahat_{\xi_i} - \thetahat_{\xi_{i-1}} \Vert<\tau
\}$, where $\tau >0$ is some
threshold value.
\end{enumerate}
By definition, $\hat{\xi}$ is the value closest to
0 such that the variation of the point estimates is smaller than some acceptable threshold.
Based on our simulations, a grid  between $\xi_1=0$  and $\xi_r = -\log(1/2)$, with
$\tau = 5\% \times \Vert \thetahat_{0} \Vert$ typically works well and choices not too far from $0$ already give
considerable stability. If a very small portion of data sub-sets are incompatible, it
may be useful to consider refinements of the grid near $\xi =0$, such as, $\xi_i = (i/n)$ , $i = 1, . . . , r$.

%
% Alternatively, since (\ref{Lcl}) is just  a type of mis-specified likelihood, model selection can be carried out by established information-theoretical
%criteria. For example, one  popular choice in the literature of composite likelihood estimation is the composite likelihood information criterion (CLIC)
% discussed in \cite{Varin05}. Here, we define $CLIC(\xi) = - 2 \ellhat(\thetahat_\xi) + \text{tr}\{\hat{H}_\xi^{-1}(\thetahat_\xi)
% \hat{K}_\xi(\thetahat_\xi)\}$, where $\hat{J}_\xi$ and $\hat{K}_\xi$ are estimates of the
% matrices $H_\xi$ and $K_\xi$ defined in Section \ref{sec:3.1}. Selection of $\xi$ can be done by computing $\text{min}_{\xi \in \Xi} \ CLIC(\xi)$, where $\Xi$
% is a grid of candidate values for $\xi$ between $0$ and $\log(m)$.

\section{Properties} \label{sec3}

\subsection{Large sample behavior of $\thetahat_\xi$ and standard errors} \label{sec:3.1}

To emphasize reliability aspects, it is helpful to distinguish between the true
process generating the data and the parametric model used for
inference. Assume that $X$ has distribution  $G(x)$, while the true distribution  for
the sub-vector $Y_j$ is denoted by $G_j(y_j)$. The density function of
$Y_j$ with respect to the dominating measure $\mu$ is denoted by $g_j(y_j)$. Let
$\{F_j(y_j;\theta), \theta \in \Theta \}$ be
a parametric family of distributions for $Y_j$ and let $f_j(y_j|\theta)$ denote the corresponding densities with respect
to $\mu$. We assume that $f_j(y_j|\theta)$ is identifiable, i.e. for
$\theta_1 \neq \theta_2$, $\mu[\{Y_j:
f_j(Y_j|\theta_1) \neq f_j(Y_j|\theta_2)\}]>0$, for all $j=1,\cdots,m$.

The composite likelihood function (\ref{comp_lik}) is correctly specified if there is a
parameter
$\theta_0 \in \Theta$ such that $f_j(y_j|\theta_0) = g_j(y_j)$ for all $1 \leq j\leq m$;
when no such  $\theta_0$ exists then (\ref{comp_lik}) is
misspecified, meaning that it contains incompatible models. The
optimal parameter, $\theta^\ast_\xi$, is defined as the minimizer of the weighted composite
Kullback-Leibler divergence
\begin{align}\label{compKL}
\theta^\ast_\xi = \underset{\theta \in \Theta}{\text{argmin}} \  E_G\left\{\log
\dfrac{g(X)}{\prod_{j=1}^m f_j(Y_j \in X|\theta)^{w_j}} \right\} =
\underset{\theta \in \Theta}{\text{argmin}} \
\sumjm w_j \ell_{j}(\theta),
\end{align}
where
$$
\ell_{j}(\theta) \equiv - E_{G_j}\ellhat_j(\theta) = - E_{G_j} \left\{ \log
f_j(Y_j|\theta) \right\}
$$
is the cross-entropy between the true distribution $G_j$ and the parametric sub-model
$f_j(\cdot|\theta)$
and  $w_j \equiv w_j(\theta)\equiv \alpha_2(\theta) \exp\{\alpha_1(\theta) \ell_j(\theta)
\} $
($j=1,\dots,m$) here denote asymptotic weights
computed as in Section \ref{computing} with
$\ellhat_j(\theta)$ replaced by $\ell_j(\theta)$. In the remainder of the paper, we assume that $\theta^\ast_\xi$ is the unique maximizer of (\ref{compKL}). 

Next, consistency and asymptotic normality of $\thetahat_\xi$ are established. We note that standard
M-estimation theory cannot be applied directly to equation (\ref{score}) because the
weights
$\{\what_j(\theta),j=1,\cdots,m\}$ in (\ref{estimator}) depend on random averages; thus some additional care
is needed to characterize the asymptotic behavior of $\thetahat_\xi$.

\begin{proposition} \label{thm1}
Assume: (C1) $\theta^\ast_\xi$ is an interior point in $\Theta$; (C2)  $\sup_{\theta \in
\Theta}
|\ellhat_j(\theta) - \ell_j(\theta)| \arrowp 0$ as
$n\rightarrow
\infty$ ($j=1,\dots,m$); and (C3) $\sup_{\theta \in \Theta} \ell_j(\theta) < \infty$
($j=1,\dots, m$). Then the maximum composite likelihood estimator $\thetahat_\xi$
converges in probability to
 $\theta^\ast_\xi$ defined in (\ref{compKL}).
\end{proposition}

\noindent A direct consequence is Fisher-consistency of $\thetahat_\xi$, i.e. under
correct composite likelihood specification the optimal target
value is $\theta^\ast_\xi = \theta_0$ for
all $\xi$. This can be seen by taking the expectation of equation (\ref{score}) with $\theta=\theta_0$:
\begin{equation}\label{eq5}
\left. E_G\left\{ \sumjm \what_j(\theta)  \uhat_j(\theta) \right\}
\right\vert_{\theta=\theta_0}= \left. E_{\what(\theta)} \left\{ \left. \sumjm
\what_j(\theta)  E_{G_j} \uhat_j(\theta)   \right\vert
\what(\theta )
\right\} \right\vert_{\theta=\theta_0} = 0,
\end{equation}
since $E_{G_j} \uhat_j(\theta_0)= 0$ if and only if $G_j(\cdot) = F_j(\cdot|\theta_0)$,
for all $1\leq j\leq m$. This means that
the estimating equation (\ref{score}) is solved by $\theta_0$ regardless of the choice of
$\xi$, since changing the latter affects only the weights $\{ \what_j(\theta)\}$, but not
the partial scores $\{ \uhat_j(\theta)\}$. Section \ref{sec:3.4}
discusses bias in the presence of incompatible sub-likelihoods.

\begin{proposition} \label{thm2}
Under conditions (C1) -- (C3) in Proposition \ref{thm1} and additional regularity
conditions given in the Appendix,  $\sqrt{n} (\thetahat_\xi -
\theta_\xi^\ast)$ converges in distribution to the $p$-variate normal $N_p(0, H_\xi^{-1} K_\xi {H_\xi}^{-1})$ as $n \rightarrow \infty$, where $H_\xi$ and
$K_\xi$ are the following $p \times p$ matrices
\begin{align} \label{asymatr}
H_\xi  =  \sumjm w^\ast_j  \left[ H_j(\thetaxi) + \alpha^\ast_1   E \{\uhat_j(\thetaxi)\}E
\{\uhat_j(\thetaxi)\}^T \right],  \ \
K_\xi =  Var\left\{ \sumjm w^\ast_j \uhat_j(\theta_\xi^\ast) \right\},
\end{align}
$H_j(\theta) = E\{\nabla_{\theta} \uhat_j(\theta)\}$, $w^\ast_j= w_j(\thetaxi)$
($j=1,\dots,m$), $\alpha^\ast_1= \alpha_1(\thetaxi)$ and  expectations are
with respect to $G$.
\end{proposition}
The random weights, $\{ \what_j(\theta) \}$, play a crucial role in determining the
asymptotic behavior of $\thetahat_\xi$.
This feature is also found in model averaging, where
parameter estimators obtained from different models, say $\hat{\mu}_S \in \mathcal{S}$,
are combined
into  a global estimator $\hat{\mu} = \sum_{s \in \mathcal{S}} \what_S \hat{\mu}_S$,
through random weights $\what_S$ \citep[Chapter 7]{claeskens08}.
The connection with model averaging is further highlighted
by the normal location example in Section \ref{examples}.
Here the random weights converge in probability to constants; thus, the
asymptotic variance takes the usual sandwich form and $H_\xi$, $K_\xi$ can be consistently
estimated analogously to
\cite{Varin&al11} with weights $
\what_j(\thetahat_\xi)$ ($j=1,\dots,m$), computed as in Section \ref{computing}.
Re-sampling
techniques such as jackknife and bootstrap may be also
used.

\subsection{Bias under incompatible models} \label{sec:3.4}

In this section, we examine the first-order properties of our estimator in the presence of incompatible models.  For clarity of exposition, in this section we
consider the
 case where
$\Theta \subseteq \mathbb{R}^1$ , but analogous  arguments can easily extended to the general case.  To represent incompatibility, we  assume
heterogeneous parameters for
the first $m^\ast$ sub-models. Particularly, let $g_j(y_j)= f(y_j|\theta_j)$, $\theta_j \in \Theta$,  ($1< j\leq m$), where
$\theta_j$, follows the drift model $\theta_j=\theta_\delta = \theta_0 + \delta$, if $j\le m^\ast$, and $\theta_j=
\theta_0$, if $m^\ast<j\le m$. In addition, we assume that the Fisher information $H_j(\theta)=E_{G} \left[\partial^2 \log f_j(X;\theta)/ \partial 
\theta^2 \right]$, $j=1,\dots, m$,  are bounded away from zero and infinity. A first-order Taylor
expansion of
$\uhat_{j}$ and
$\ellhat_{j}$ in (\ref{score}) about $\theta_j$ under suitable
regularity conditions gives
\begin{align*}
0 =\sumjm \exp\{\alpha_1(\thetahat_\xi) \ellhat_{j}(\thetahat_\xi) \} u_{nj}(\thetahat_\xi)
\approx \sum_{j=1}^m \exp\{\alpha_1^\ast
\ell_{j}(\theta_j)\}
\left[
\thetahat_\xi - \theta_0 + \theta_0 - \theta_j \right] H_j(\theta_j).
\end{align*}
Re-arranging the above expression  leads to the following approximation for the bias
\begin{align}\label{Bias}
\thetahat_\xi - \theta_0  \approx
\frac{m^\ast\delta \exp\{\alpha_1^\ast \ell_{1}(\theta_\delta)\}H_1(\theta_\delta)}{m^\ast\exp\{\alpha_1^\ast \ell_{1}(\theta_\delta)\}
H_1(\theta_\delta)+(m-m^\ast)\exp\{\alpha_1^\ast \ell_{n1}(\theta_0)\}H_1(\theta_0)}\nonumber
=\frac{\delta}{1+C(\theta_0, \delta)},
\end{align}
where
\begin{align*}
C(\theta_0, \delta)= \frac{(m-m^\ast)H_1(\theta_0)}{m^\ast H_1(\theta_\delta)}\exp\{\alpha_1^\ast(\ell_{1}(\theta_0)-\ell_{1}(\theta_\delta))\}\ge
\frac{c_1}{c_2}\left( \dfrac{m}{m^\ast} - 1\right) \exp\{\alpha_1^\ast(\ell_{1}(\theta_0)-\ell_{1}(\theta_\delta))\}.
\end{align*}
Therefore, an approximate upper bound
to the bias, $\vert \thetahat_\xi - \theta_0 \vert$, is
\begin{equation} \label{upper}
\text{Max-Bias}(\thetahat_\xi|\delta) \equiv
\dfrac{\vert \delta \vert}{ 1 +  \dfrac{c_1}{c_2}\left( \dfrac{m}{m^\ast} - 1 \right)  \exp\left\{
-\dfrac{\alpha_1^\ast  \delta^2
H_1(\theta_0)}{2}  \right\} },
\end{equation}
which is regarded as  the worst-case bias under incompatible models. Clearly, when $\xi= 0$ (equivalently, $\alpha^\ast_1=0$), the worst-case bias grows
linearly in $\delta$. When $\xi>0$, $\text{Max-Bias}(\thetahat_\xi| \delta)$ is bounded and the estimator $\thetahat_\xi$ achieves bias control. Particularly,
if $\delta=0$ and all the models are compatible, then $\text{Max-Bias}(\thetahat_\xi| \delta) = 0$. If $\delta$ is large, since the denominator in
(\ref{upper}) dominates the numerator, the maximal bias decreases quickly to 0.

A second-order Taylor expansion of $\uhat_{j}$ and $\ellhat_{j}$ in (\ref{score}) about $\theta_j$ (not shown here) can be used to derive an upper bound for 
the mean squared error. Analogously to (\ref{upper}), when $\xi= 0$ (equivalently, 
$\alpha^\ast_1=0$), the worst-case mean squared error grows quadratically in $\delta$. When $\xi>0$, the maximal mean squared error is bounded, meaning that 
the estimator $\thetahat_\xi$ achieves both bias and variance control. This theoretical understanding is confirmed by the numerical simulations in Section 4.

\begin{figure}[h]
\centering
\begin{tabular}{ccc}
\includegraphics[scale=0.36]{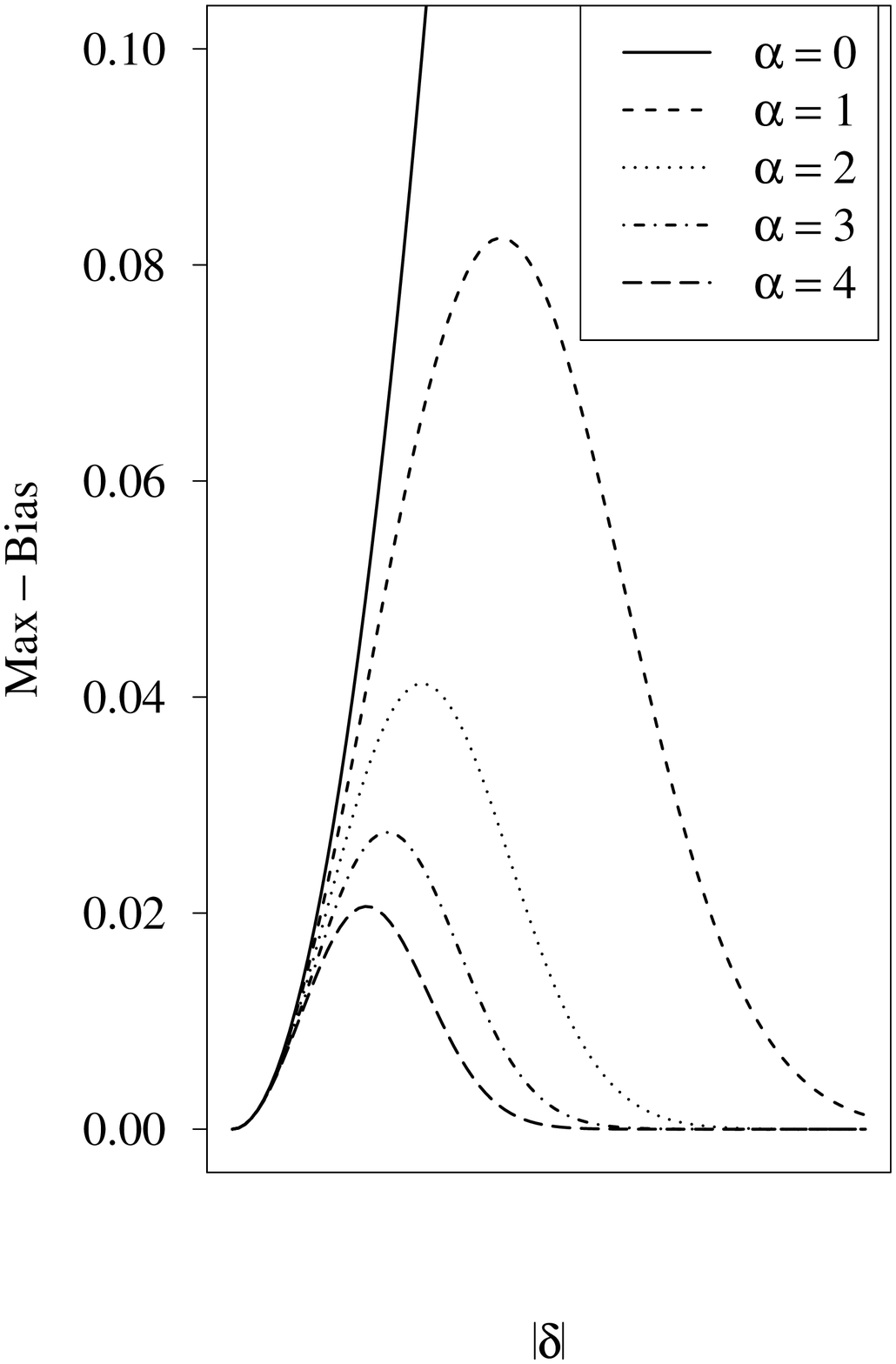}
&\includegraphics[scale=0.36]{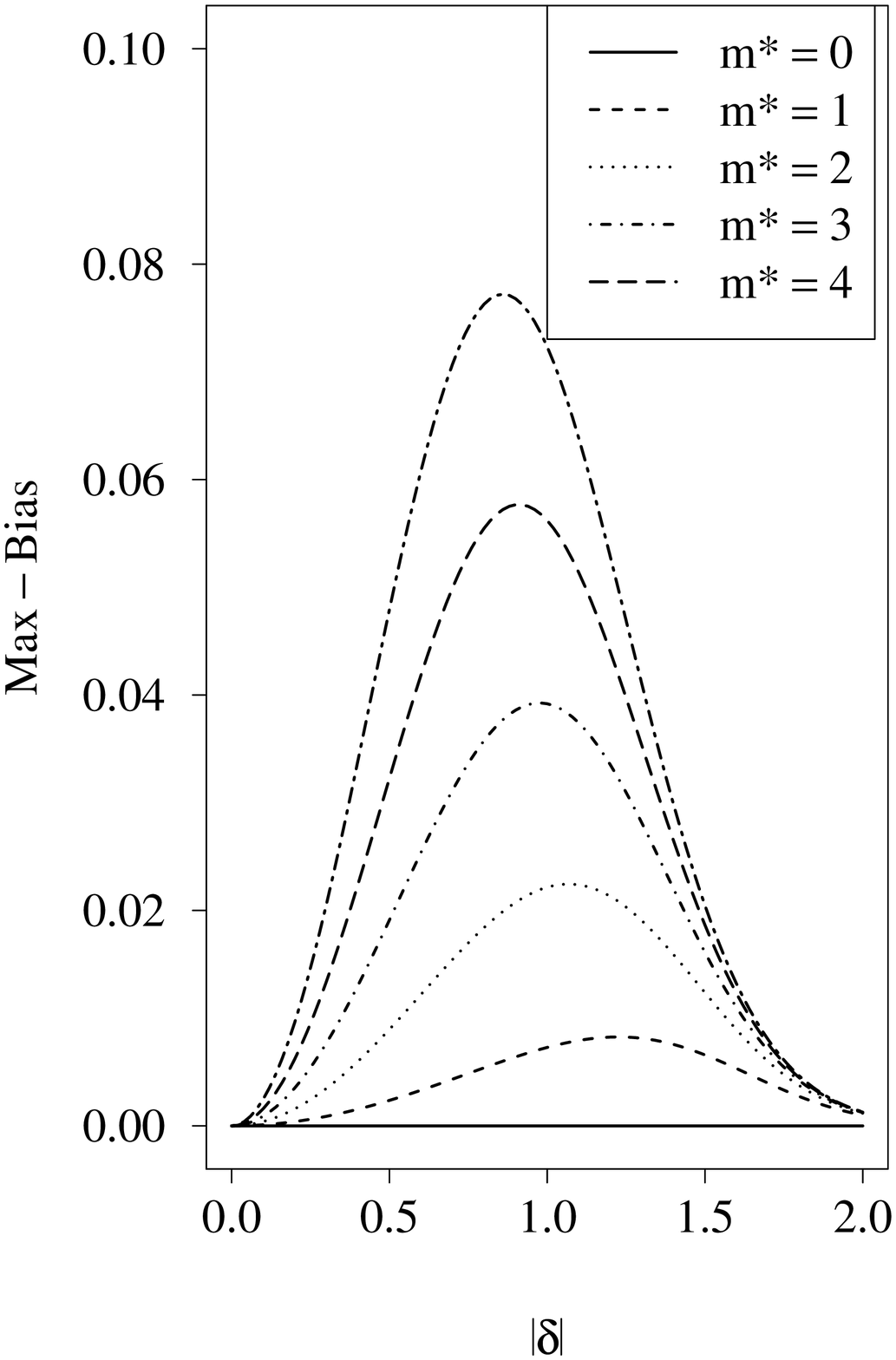}
\end{tabular}
\caption{Worst-case bias for the multivariate normal location model $X\sim N_{10}(\theta,I)$  with $\theta_j = \theta_0 + \delta$ where  $\delta = 0$, if
$1 \leq j \leq m^\ast$, and  $\theta_j = \theta_0$, if $m^\ast < j \leq 10$ .  Left: the curves
correspond to different values of the constant $\alpha_1^\ast$ described in Sections \ref{sec:3.1} and \ref{sec:3.4}
($\alpha_1^\ast=0,1,2,3,4$,  and $m^\ast =1$). Right:  the curves correspond to increasing number of incompatible models, $m^\ast$, ranging from $0$
(horizontal solid line)  to 4 ($\alpha_1^\ast=1$).}
\label{figBias}
\end{figure}

As an illustration, Figure \ref{figBias} shows the maximal bias for the multivariate normal
model $X\sim N_{m}(\theta, I)$ with $\theta_j = \theta_0 + \delta$ where  $\delta = 0$ if $j=1,\dots, m^\ast$, and  $\theta_j = \theta_0$, if
$m^\ast \leq j \leq m$. Clearly, the classic
estimator with  equal weights  ($\xi=0$)  is very risky for this model, since the maximal bias can be potentially very large. This undesirable
behavior can be easily avoided by setting $\xi>0$. Thus, if the degree of incompatibility is strong ( $|\delta| \rightarrow \infty$), the worst-case
bias approaches zero. For intermediate cases where $|\delta|<\infty$ the bias remains bounded and can be controlled  by tuning
$\xi$.

\section{Examples}  \label{examples}

\subsection{Example 1: Estimation of correlation} Suppose the random vector $(X_1, X_2,
X_3,
X_4, X_5)^T$ follows a
multivariate normal distribution  with zero mean vector, unit variances and covariances
$Cov(X_1, X_k) = \rho_0 / \sqrt{\varepsilon}$ if $2 \leq k \leq 5$, for some
$\varepsilon \geq 1$, and $Cov(X_j, X_k) = \rho_0$ otherwise.
If we model $X$ as a multivariate normal with zero mean vector and all correlations equal,
then the model is clearly misspecified and the maximum likelihood estimator is not
consistent for $\rho_0$.

When constructing a composite likelihood function we only need pair-wise lower-dimensional likelihoods, since the marginal univariate sub-likelihoods
do not contain information on $\rho_0$.  Therefore the correlation estimator $\hat{\rho}_\xi$ is
obtained as described in Section \ref{sec2} by maximizing the pairwise likelihood
\begin{eqnarray}\label{corr_lik}
\ellhat(\rho|w) = \sum_{j>k} w_{jk} \ellhat_{jk}(\rho) \equiv \sum_{j>k} w_{jk} \left\{ -
\frac{n}{2} \log (1-\rho^2) - \frac{(SS_{jj}
+ SS_{kk})}{2(1-\rho^2)} + \frac{\rho  SS_{jk}}{1-\rho^2} \right\},
\end{eqnarray}
where  $SS_{jj} = \sumin (\xsubi_j)^2$ and $SS_{jk}= \sumin \xsubi_j \xsubi_k$. Note that
(\ref{corr_lik}) refers to combining bi-variate
normal models with zero mean and covariance given by $2 \times 2$ matrices with
diagonal elements equal to 1 and off-diagonal elements equal to $\rho$.  Therefore
$\hat{\rho}_\xi$
will be consistent for $\rho_0$ only if  $\what_{12}=\cdots \what_{15}=0$.

In Table \ref{table1}, we show the finite-sample  bias and variance of the D-McLE for different values of $\xi$. As a comparison,
we report results for the MLE and the usual McLE with uniform weights corresponds to the column with $\xi=0$. When all the sub-likelihoods are compatible
($\varepsilon = 1$), not surprisingly the MLE has the best performance in terms of variance. For the D-McLE, however, both bias and variance do not increase
much as long as $\xi$ is not too far from $0$.  In the presence of incompatible sub-models ($\varepsilon=3,5$),  the bias for the MLE and D-McLE with uniform
weights ($\xi=0$) is very large compared to the D-McLE with $\xi>0$. For example, when $\varepsilon=3$, the bias of the D-McLE is negligible when
$\xi=0.2$. In addition to bias control of D-McLE, we note also that our procedure also achieves variance reduction
when $\xi>0$ and $n$ is small. These results suggest
that by setting $\xi$ slightly above zero (e.g., 0.1, 0.2, or 0.3) already gives substantial stability and reduce the mean squared error of the corresponding
estimator, $\thetahat_\xi$.

\begin{table}[ht]
\begin{tabular}{c|cccccccccccccccc}
 & MLE     & &\multicolumn{11}{c}{D-McLE($\xi$)}  \\
$\varepsilon$  &         &$\xi$=& 0.0 & 0.1 & 0.2 & 0.3 &
0.4& 0.5 & 0.6 & 0.7 & 0.8 & 0.9 & 1\\
\hline
  &     &\multicolumn{11}{c}{Bias$^2 \times 100$}  \\
 1      & 0.00  & & 0.00 & 0.06 & 0.14 & 0.21
& 0.27 & 0.37 & 0.43 & 0.52 &
0.62 & 0.68 & 0.78\\
 3      & 2.43   & & 1.75 & 0.12 & 0.00 & 0.05
& 0.12 & 0.20 & 0.26 & 0.36 &
0.43 & 0.50 & 0.58\\
 5     & 6.32   && 4.53 & 0.46 & 0.00 & 0.05
& 0.11 & 0.19 & 0.27 & 0.34 &
0.42 & 0.51 & 0.57 \\
  &   &\multicolumn{11}{c}{Var$\times 100$}  \\
 1 &      0.10 && 0.12 & 0.12 & 0.12 & 0.13 &
0.14 & 0.13 & 0.13 & 0.14 &
0.14 & 0.14 & 0.15 \\
3 &      0.18 && 0.20 & 0.14 & 0.13 & 0.13 &
0.14 & 0.13 & 0.14 & 0.14 &
0.15 & 0.16 & 0.16 \\
 5 &      0.18 && 0.22 & 0.15 & 0.13 & 0.13 &
0.14 & 0.14 & 0.14 & 0.15 &
0.15 & 0.16 & 0.17
\end{tabular}
\label{table1}
\caption{Bias and variance for pairwise likelihood estimation of the correlation model
$N_5(0, \Sigma)$ with unit variances and $Cov(X_1, X_k) = \rho_0 /
\sqrt{\varepsilon}$ if
$2 \leq k \leq 5$, and $Cov(X_j, X_k) = \rho_0$ otherwise, with $\rho_0=1/2$ and
$\varepsilon=1,3,5$ ($\varepsilon=1$ corresponds to the correctly specified model). The
columns refer to maximum likelihood estimator (MLE) and the discriminative composite likelihood
estimator (D-McLE) with $\xi$ ranging from 0
to 1 ($\xi=0$ implies uniform weights). Results are based on $10^4$ Monte Carlo samples of
size $n=50$.}
\end{table}

\begin{figure}[h]
\centering
\begin{tabular}{cc}
\includegraphics[scale=0.3]{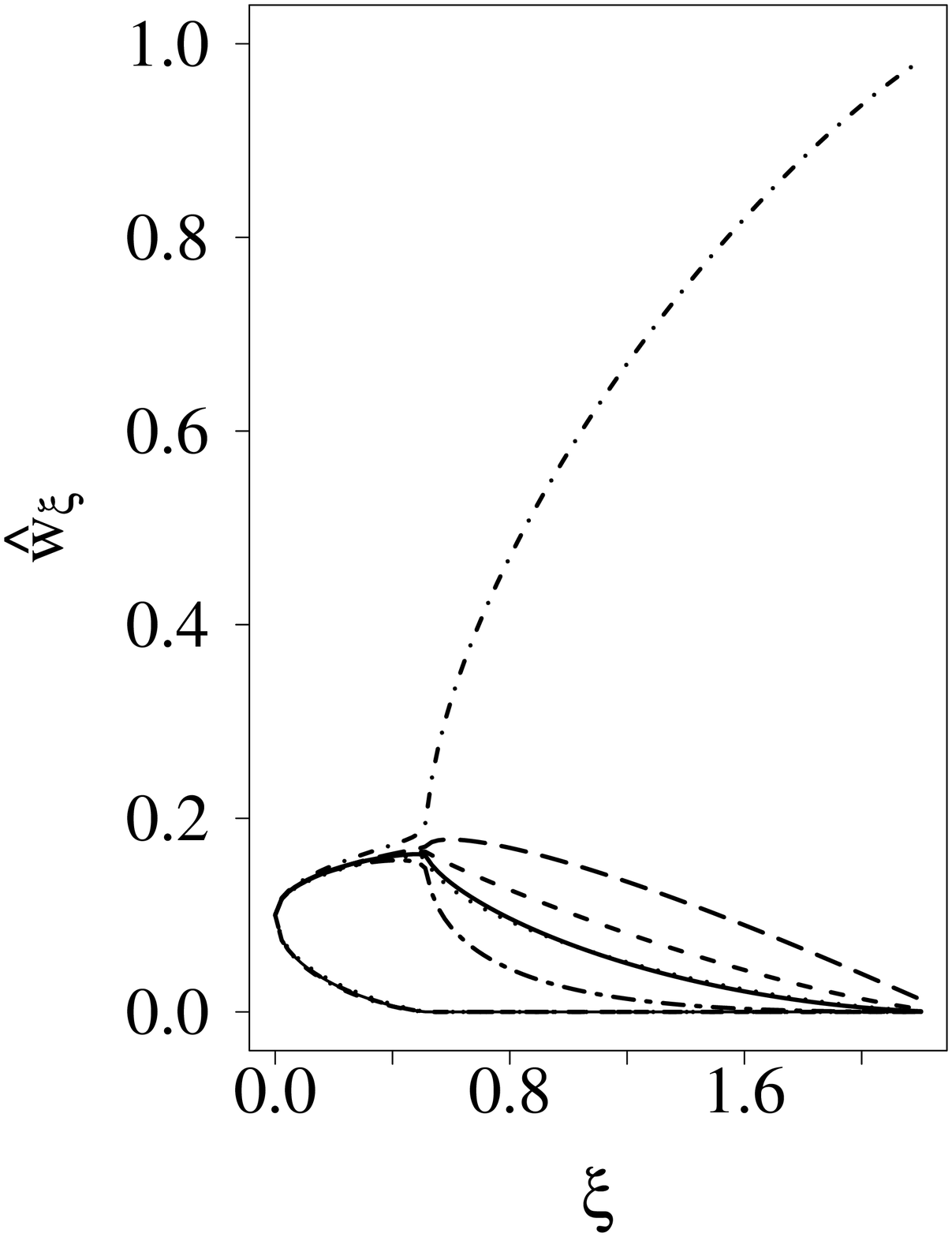}
&\includegraphics[scale=0.3]{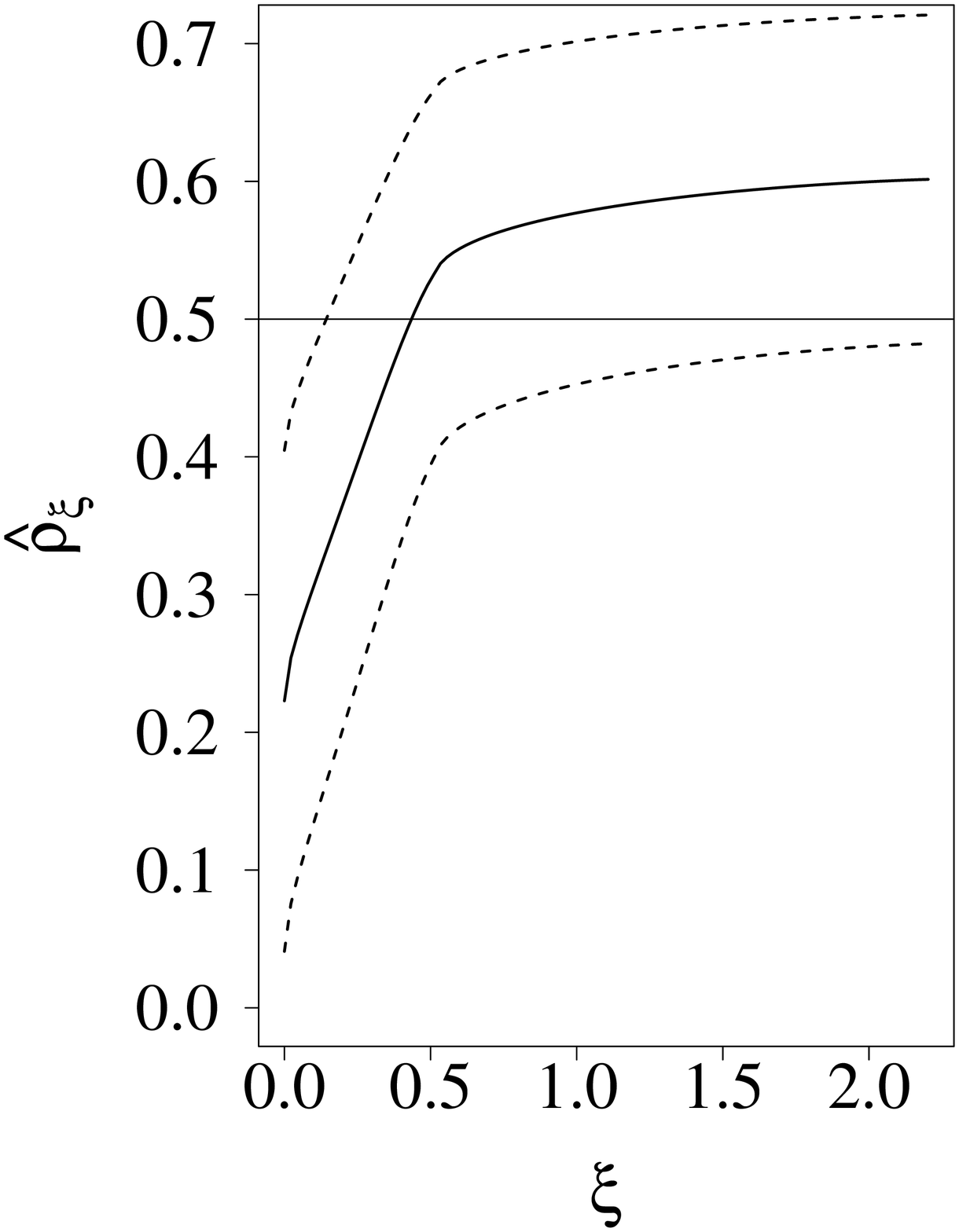}
\end{tabular}
\caption{Estimation of the correlation  model $N_5(0,
\Sigma)$ with unit variances and $Cov(X_1, X_k) = \rho_0 / \sqrt{5}$ ($2 \leq k \leq
5$),
and
$Cov(X_j, X_k) = \rho_0$ ($j\neq k \neq 1$), with true parameter $\rho_0=1/2$. Left:
Importance
profile paths for the partial likelihood components based on the estimated weights,
$\what_\xi$. Right: estimated correlation coefficient (horizontal is the
true value $\rho_0 = 0.5$). Illustration based on 50 observations.}
\label{fig1}
\end{figure}

Figure \ref{fig1} illustrates the profile plot
(left) and parameter estimates
(right) for a sample of $n=50$ observations. When $\xi=0$, the estimator is unreliable with estimates between  $
\rho_0/\sqrt{\varepsilon} = 0.5/\sqrt{5}\approx 0.22$ and
$\rho_0=0.5$. When $\xi$ moves away from zero, the importance profile shows two distinct
groups of sub-likelihoods, with the
four overlapping paths at the bottom corresponding
to  misspecified sub-likelihoods. When $\xi= 0.51$, the estimator $\hat{\rho}_\xi$ exploits correctly the information from
the compatible sub-likelihoods  and gives estimates close to the true value $\rho_0=1/2$. Finally, as $\xi
\rightarrow \log(10)$, a single partial likelihoods tends to dominate
the others, but much of the information from the other useful data pairs is ignored. Therefore the composite estimate at $\xi=\log(10)$
is inferior to that at $\xi=0.51$, in terms of accuracy.

\subsection{Example 2: Location of heterogeneous normal variates} \label{ex:norm}
Let $(X_1,
\dots, X_m)$ be independent normal variables with common mean $E(X_j) = \mu_0$ ($1
\leq j\leq m)$ and heterogeneous variances $Var(X_j) = \sigma_{0,j}^2$
($1 \leq j \leq m$). This is the basic
meta-analysis model where a weighted average of a series of study estimates, say
$\{ \overline{X}_j \}$, is  combined to obtain a more precise estimate
for $\mu_0$. The inverse of the estimates' variance,
$1/\sigma^2_j$, is the optimal study weight ensuring minimum variance of the
combined
estimate. All the parameter information is contained in the marginal models, so the
following
negative one-wise composite likelihood function is minimized:
\begin{align}\label{normallik}
- 2 \ellhat(\mu, \sigma_1, \dots, \sigma_m| w)  = \sumjm w_j\left\{ \log \sigma_j^2 +
\dfrac{1}{n}\sumin \dfrac{(X^{(i)}_{j} - \mu)^2}{\sigma_j^2} \right\}.
\end{align}
and the profiled composite likelihood estimators are
\begin{equation*}
\hat{\mu}(w) \equiv \sumjm w_j \overline{X}_j \equiv \sumjm  \dfrac{w_j}{n} \sumin
X^{(i)}_j, \ \
\hat{\sigma}^2_j(w)\equiv \dfrac{1}{n}\sumin \{X^{(i)}_{j} - \hat{\mu}(w)\}^2, \ \ j=1,\dots,m.
\end{equation*}
Replacing
$\mu = \hat{\mu}(w)$ and
$\sigma_j = \hat{\sigma}_j(w)$ in (\ref{normallik})
gives $ \sumjm w_j \log
{\hat{\sigma}^2_j(w)}$, which is then minimized subject to the constraints  $D_{KL}(w) =
\xi$ and $\sumjm w_j =1$.

The resulting location estimator, say
$\hat{\mu}_\xi$, solves the fixed-point equation
\begin{equation}\label{fixed-point}
\mu = \sumjm\hat{w}_j(\mu) \overline{X}_j = \dfrac{\sumjm \overline{X}_j  \{
\sumin (X^{(i)}_{j}
- \mu)^2 \}^{-\hat{\alpha}_1}  }{\sumjm \{ \sumin (X^{(i)}_{j}
- \mu)^2\}^{-\hat{\alpha}_1}},
\end{equation}
where $\hat{\alpha}_1 >0$ is computed as in (\ref{weights}) for a given  $\xi \geq 0$, and the variance estimators are
$\hat{\sigma}^2_{\xi,j} = n^{-1} \sumin (X^{(i)}_{j} - \hat{\mu}_\xi)^2$ ($j=1,\dots,m)$.

The degree of incompatibility of models is very strong, then the estimator $\hat{\mu}_\xi$ is nearly as good as the estimator obtained by ignoring the
corresponding data sub-sets. If all the models are compatible, $\hat{\mu}_\xi$ still performs well in terms of accuracy. Particularly, if all the
partial likelihoods are correctly specified, then $E(\hat{\mu}_\xi) = \mu_0$. If $\overline{X}_j$ ($1\leq j\leq m^\ast$) are far away from $\mu_0$,
then finding $\hat{\mu}_{\xi}$ is approximately equivalent to solving (\ref{fixed-point}) with
$\what_{j}(\mu) = 0$, if $j \leq m^\ast$.

Table \ref{table2} shows bias and  variance for
$\hat{\mu}_\xi$ under
correctly specified and misspecified sub-likelihoods. The usual McLE with uniform weights corresponds to the column with $\xi=0$.
For comparison purposes, we also show the maximum likelihood estimator with weights $
w_{mle,j} \propto 1/S_j^2$, where $S^2_j$ is the sample
standard deviation for the $j$th variable. The results
correspond
to the location model with $\sigma_{0,j} = 1/j$ ($j=1,\dots, 10$)
and misspecification introduced by the location shift $\mu_j = \mu_0 + 1$, $j=1,2$. When all the sub-likelihoods are compatible
($m^\ast = 0$),  the MLE has the best performance, but the D-McLE with $\xi = 0.1$ doing comparably well.  In the presence of two incompatible sub-models
($m^\ast = 2$),  the bias for the MLE and D-McLE with uniform  weights ($\xi=0$) is  large compared to the D-McLE with $\xi>0$. The bias is quite small when
$\xi = 0.3$. The variance of D-McLE for $0<\xi \leq 0.3$ is also quite small compared to the McLE with uniform weights; interestingly in a few cases the
variance is smaller than that of the MLE. This confirms the behavior observed in other numerical examples and in the derivations given in Section
\ref{sec:3.4}. Across a number of other simulation settings, we found that $\xi$ slightly larger than zero  gives estimators with negligible bias  and
relatively small mean squared errors.

\begin{table}[htp]
\centering
\begin{tabular}{cc|cccccccccccccc}
             &   & MLE&  &  \multicolumn{8}{c}{D-McLE($\xi$)} \\
$n$  &          $m^\ast$& & $\xi=$  &  0.0 &0.1& 0.2 &0.3 &
0.4
&  0.5 &0.6& 0.7 \\
\hline
&          & & \multicolumn{8}{c}{Bias$^2 \times 1000$} \\
$10$          &0 & 0.00&&  0.01& 0.00& 0.01& 0.07&
0.14& 0.19& 0.21& 0.22\\
              &2 & 1.35&& 36.32& 1.27& 0.16& 0.09&
0.22& 0.26& 0.59& 1.79   \\
$100$         &0 & 0.00&&  0.00& 0.00& 0.00& 0.00&
0.00& 0.00& 0.00& 0.00\\
              &2 & 2.53&& 39.47& 1.14& 0.21& 0.04&
0.00& 0.00& 0.00& 0.01     \\
&         & & \multicolumn{8}{c}{Var$\times 1000$} \\
 $10$         &0 & 3.51&& 5.03& 4.08& 6.69&  9.55&
11.01& 11.89& 12.49& 12.90 \\
              &2 & 9.50&& 6.66& 5.06& 6.93& 10.23&
15.14& 17.58& 18.70& 21.67  \\
$100$         &0 & 0.41&& 0.65& 0.43& 0.47&
0.53& 0.59& 0.65& 0.71& 0.76   \\
              &2 & 0.53&& 0.73& 0.48& 0.51&  0.53&
0.55& 0.57& 0.59& 0.61
\end{tabular}
\label{table2}

\caption{Bias and variance for location estimates of $X\sim N_{10}(\mu_0,
\Sigma_0)$, where $\Sigma_0 = \text{diag}(1,
1/2, \dots, 1/10)$,  with and without incompatible models ($m^\ast=0,2$, respectively).
The columns correspond to the maximum likelihood estimator (MLE) with
weights proportional to $\{1/S_j^{2}\}$, where $S^2_j$ is the sample standard deviation
for the $j$th variable, and the composite likelihood estimator with $\xi$ between 0 and 0.7
(D-McLE). For $m^\ast=2$, misspecification is
introduced as $\mu_j = \mu_0 + 1$, $j=1,2$. Results based on $10^4$ Monte Carlo samples
of sizes $n=10,100$.}
\end{table}

\section{Multivariate models for spatial extremes: application to the Tasmanian rainfall
data} \label{sec5}

Max-stable processes have emerged as a useful representation of
extreme environmental occurrences such as hurricanes, floods and storms
\citep{davison2012}. However, their estimation poses significant challenges, since they lack of a general multivariate density expression. A
well studied case  is
the Gaussian max-stable process defined as
$Z(s) \equiv \max_{i\geq 1} \{V_i f(U_i-s)\}$, where $\{V_i, U_i\}$ is a Poisson process
on $(0,\infty] \times \mathbb{R}^2$, with intensity measure $\nu(ds)\times u^{-2} du$,
and $f$ is the bivariate normal distribution with zero mean and covariance
$\Sigma$ \citep{Smith90}. The process $Z$ has
unit Frech\'{e}t margins with distribution function $F(z)= \exp(-1/z)$, $z>0$.
\cite{Smith90} interprets $Z$ as extreme environmental episodes, such as storms,
where $V$, $U$, and $f$ are the storm magnitude, center,  and shape, respectively.

Next, we apply the D-McLE to estimate the extreme covariance parameter $\Sigma$ in the context
of the Tasmania rainfall data described below. For a finite set of spatially-referenced indexes, $s_1, \dots, s_d \in \mathbb{R}^2$, the
joint distribution of the random vector $Z(s_1), \dots, Z(s_d)$ has no
analytical representation for $d>2$.  \cite{Padoan10} give a closed-form
expression for the bivariate density and propose estimation based on the pairwise
likelihood function.  Given $n$
observations on $d$ locations, $z^{(i)}_1,\dots,
z^{(i)}_d$, $(i=1,\dots,n)$, the weighted pairwise likelihood function
obtained by considering all  $m(m-1)/2$  location pairs is
$$
\ellhat(\Sigma|w)= \sum_{j=1}^{m-1} \sum_{k=j+1}^{m} w_{jk} \sumin \log f_{Z_j
Z_k}\left(\left.
z^{(i)}_j , z^{(i)}_k \right\vert \Sigma \right), $$
where $f_{Z_j Z_k}$ is the bivariate density
\begin{align}
& f_{Z_j Z_k}(z_j, z_k| \Sigma) =   \exp\left[  \dfrac{ \Phi\{ g_1(h) \} }{z_j}  -
\dfrac{
\Phi\{
g_2(h) \} }{z_k}  \right] \times
\left\{ \left[
\dfrac{  g_2(h) \varphi\{g_1(h)\} }{a(h)^2 x^2_jz_k}  -  \dfrac{ g_1(h) \varphi\{g_2(h)\}
}{a(h)^2 z_jx^2_k}
\right]\right. \notag \\
 & +
\left.
\left[
\dfrac{ \Phi\{g_1(h)\} }{x^2_j}  +  \dfrac{  \varphi\{g_1(h)\}}{a(h)^2 x^2_j}
- \dfrac{  \varphi\{g_2(h)\}}{a(h)^2 z_j z_k}
\right]
\left[
\dfrac{ \Phi\{g_2(h)\} }{x^2_k}  +  \dfrac{  \varphi\{g_2(h)\}}{a(h)^2 x^2_k}
- \dfrac{  \varphi\{g_1(h)\}}{a(h)^2 z_j z_k}\right]
\right\} \label{gauss}.
\end{align}
In the above expression, $\Phi$ and $\varphi$ are the standard normal probability and density
functions, respectively;  $h = (s_j - s_k)$, $a(h)= (h^T \Sigma h)^{-1/2}$; $g_1(h)=
a(h)/2 + \log(x_j/z_k)/a(h)$; and $g_2(h) = a(h) - g_1(h)$. For
fixed $h$, the extremal dependence behaviour is determined by $\Sigma$, which is
therefore the main interest for inference. Since the above model
requires unit Frech\'{e}t margins, the observed margins, $y_j$,
are transformed in unit Frech\'{e}t by the transformation
$
y_j = g_j(y_j) \equiv \left[ 1 + \zeta_j {\{ y_j - \mu_j\}}/{\gamma_j}
\right]_+
$,
where $u_+ = \max(0,u)$ and $\mu_j$, $\gamma_j$ and $\zeta_j$ are location, scale
and shape parameters obtained from the empirical distribution.

\begin{figure}[h]
\centering
\begin{tabular}{ccc}
\includegraphics[scale=0.3]{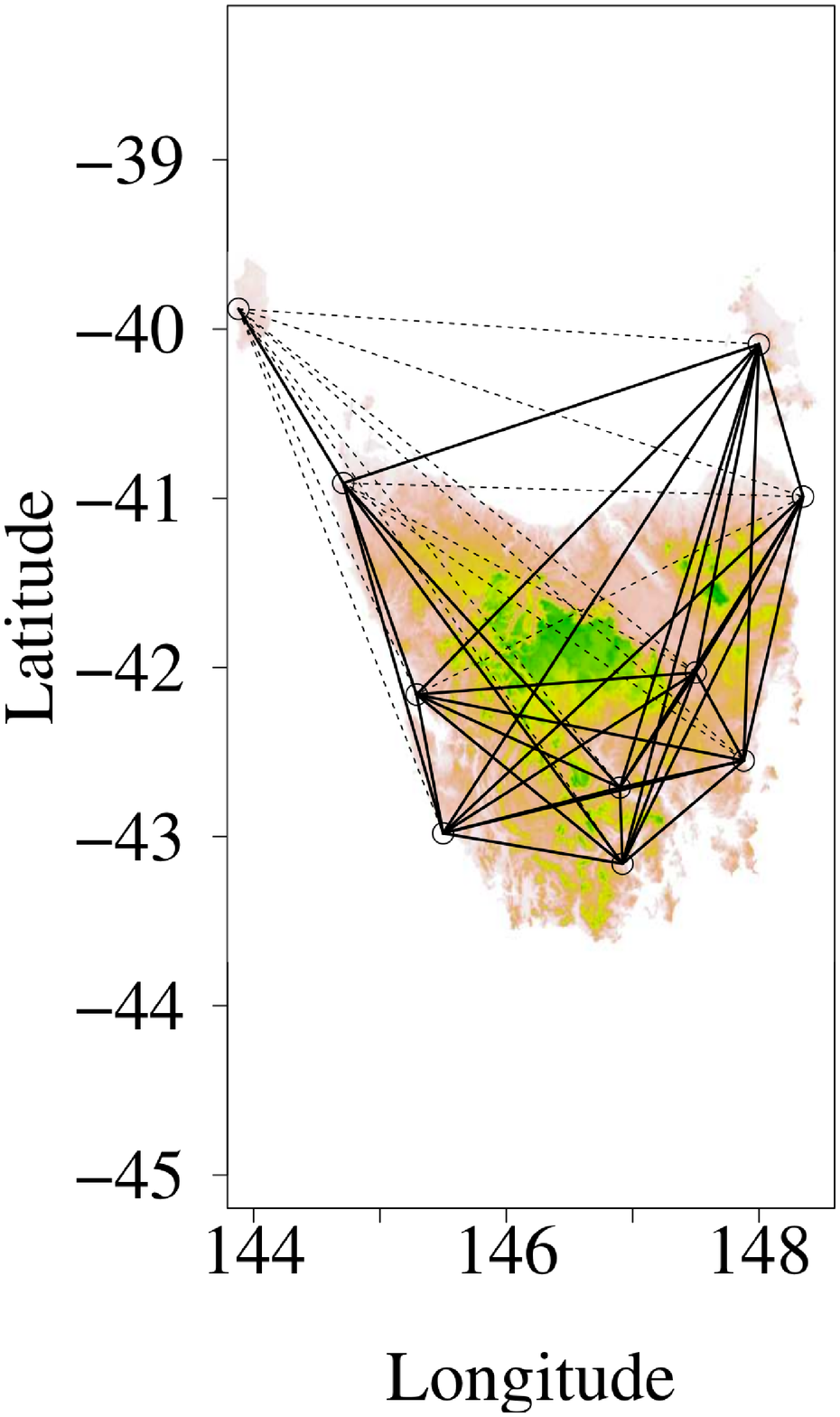}
& \includegraphics[scale=0.3]{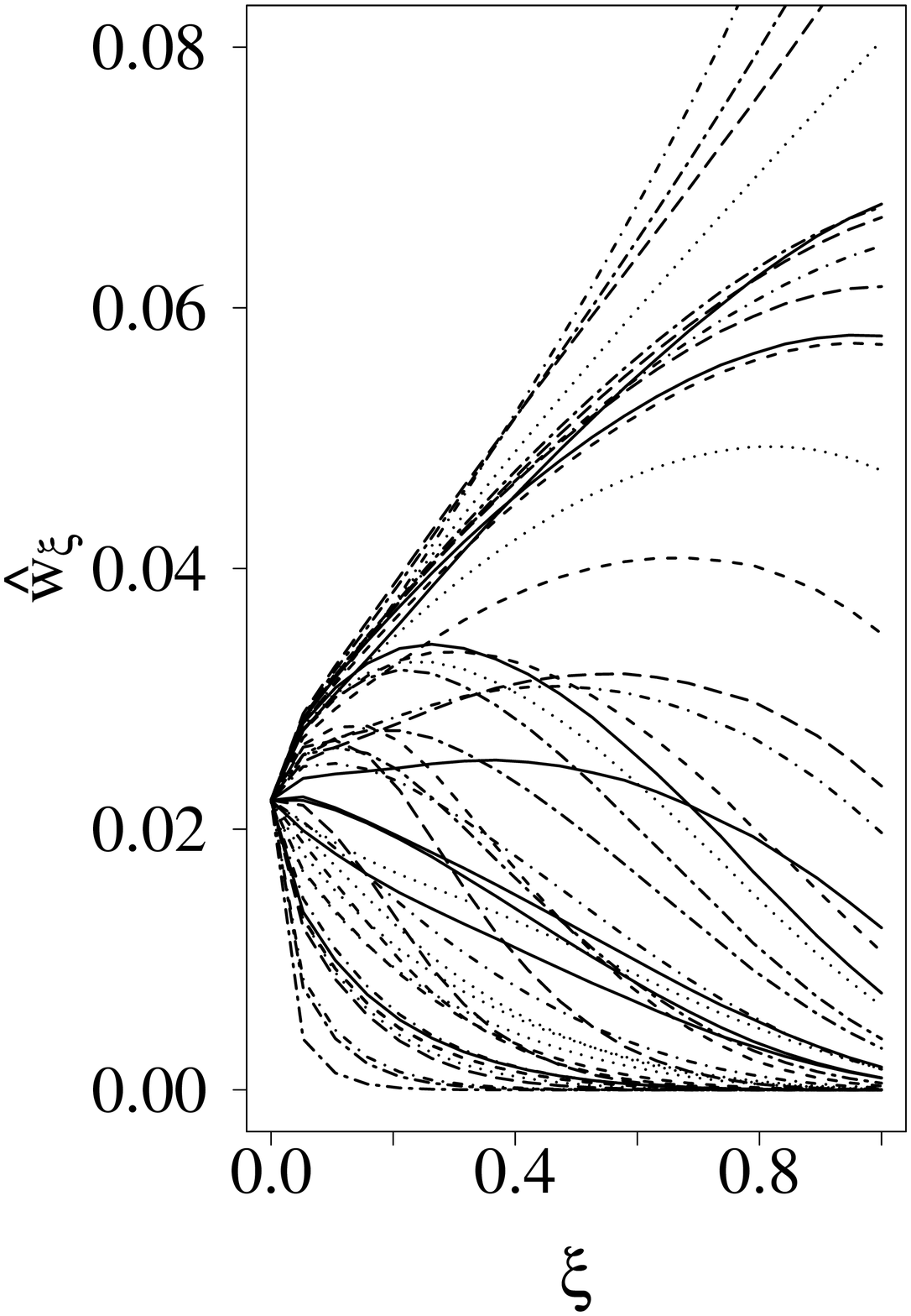}
\end{tabular}
\caption{Left: Tasmania elevation map with location of
the weather gauging stations. The dashed edges denote fitted weights  $w_{nj}$ (computed as described in Section \ref{computing})  smaller than the first
quartile for the weights ($\approx 0.0065$) when $\xi=0.3$.  Right: compatibility profile plots for $\xi$ between 0 and 1.}
\label{fig2}
\end{figure}

We consider a data set of 20 yearly rainfall maxima recorded at 10 gauging stations from 1995 to 2014 in the Australian state of
Tasmania corresponding to the following
locations, also shown in Figure \ref{fig2}: Bushy Park, Ross, King Island, Eddistone Point, Geeveston, Strahan, Flinders Island, Marrawah, Rocky Point, Orford
(source:
\texttt{http://wwwc.bom.gov.au/tas/}). The max-stable
Gaussian model is then fitted using a pair-wise likelihood function including all $m = {{10}\choose{2}} = 45$ pairs of locations. We compute estimates
$\hat{\Sigma}_\xi$
for different choices of $\xi$ ranging from $0$ to $\log(45)$. Figure \ref{fig2} (left) shows the a map of Tasmania with the 10 stations locations, and the
edges denote fitted weights, $\hat{w}_{nj}$ corresponding to $\xi = 0.3$ (dashed lines represent weights smaller than the first quartile of
 fitted weights). Figure \ref{fig2} (right) shows CPP plots for the weights.  We note that pairwise likelihoods involving the King Island station
(located at coordinates  -39.88 , 143.88 on the map) exhibit a  very weak degree of compatibility compared to locations in the southern and eastern part of
the island. This suggest a different pattern for the precipitations for King Island in relation to the rest of the stations; thus  pair-wise
sub-models involving such a station should be further inspected and possibly revised.

\begin{figure}[h]
\centering
\begin{tabular}{ccc}
\includegraphics[scale=0.26]{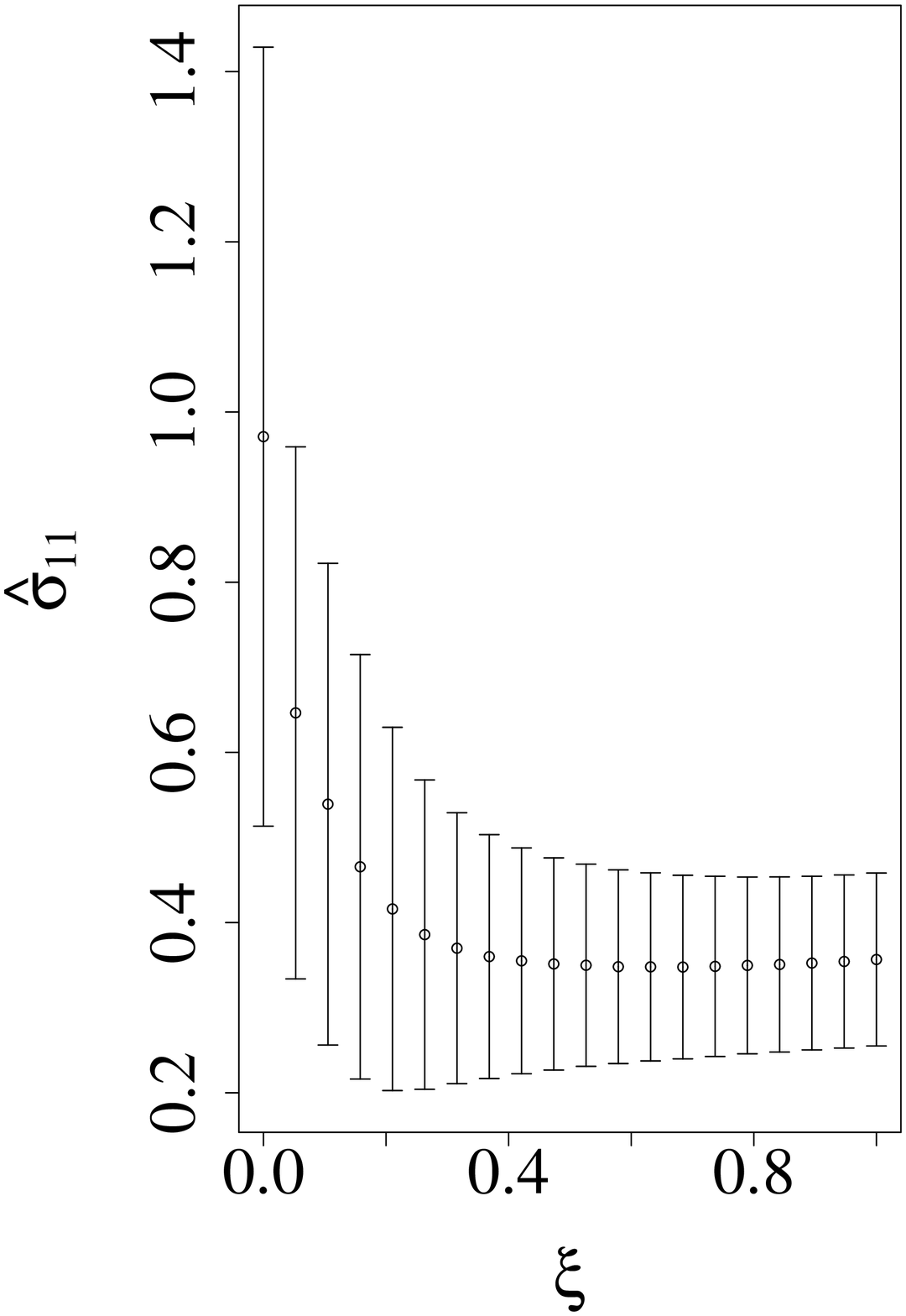}
&\includegraphics[scale=0.26]{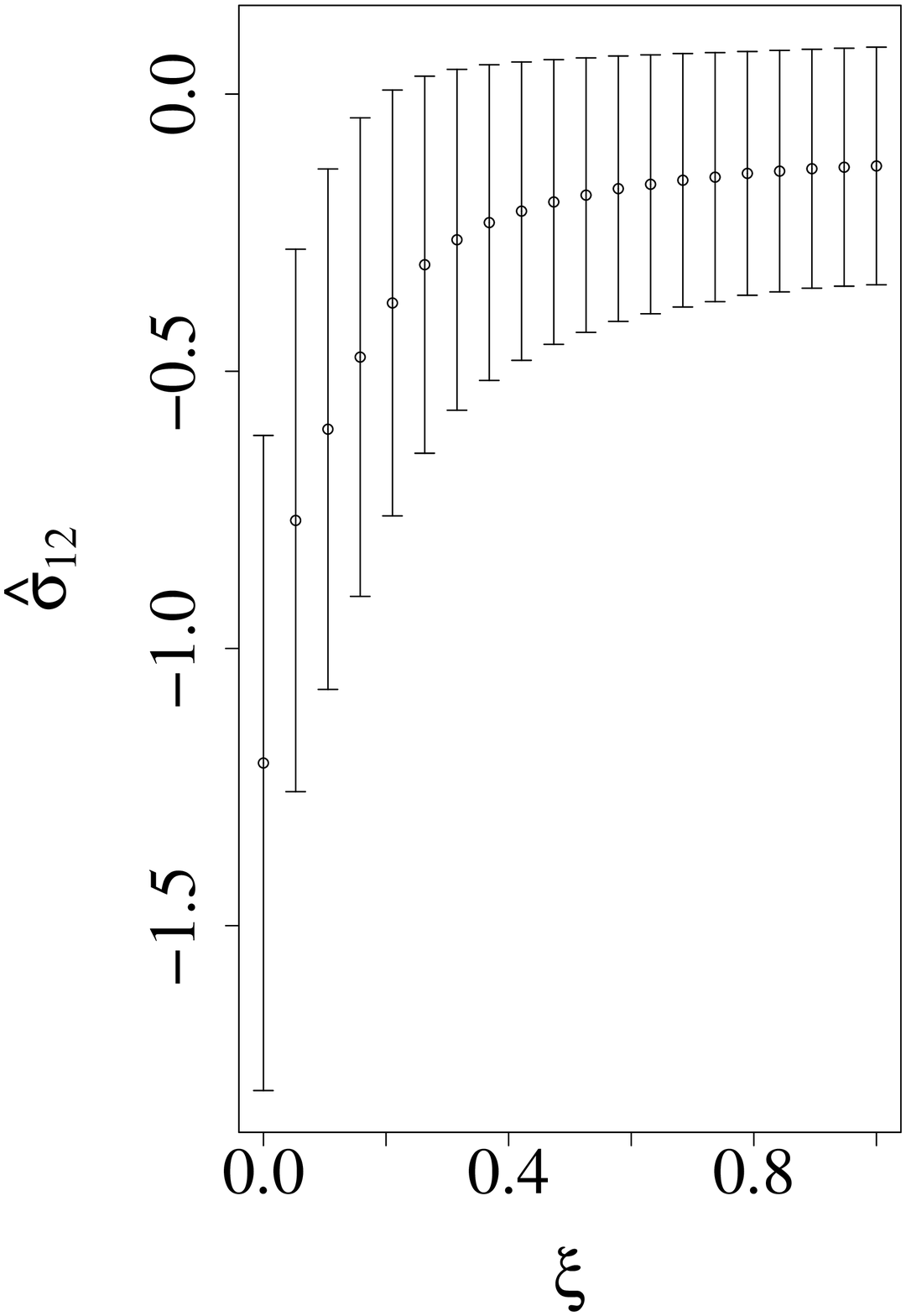}
&\includegraphics[scale=0.26]{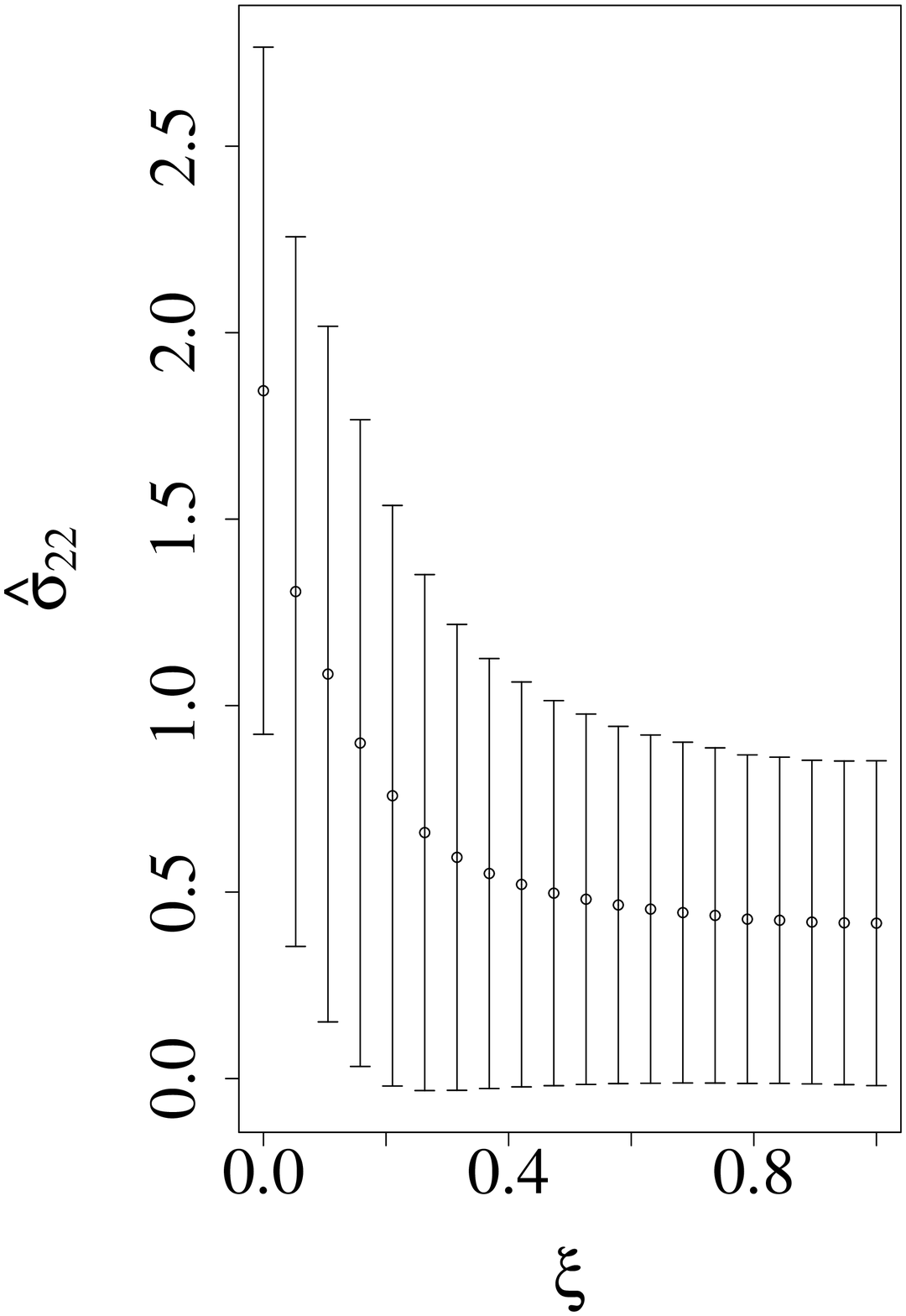}
\end{tabular}
\caption{Estimation of the Gaussian max-stable model for the Tasmania rainfall data. Estimates of $\sigma_{11}, \sigma_{12},\sigma_{22}$ for $\xi$ ranging from
0 to 1. Vertical bars represent 95\% confidence intervals based on standard errors from the asymptotic distribution of the D-McLE.}
\label{fig3}
\end{figure}

Figure \ref{fig3} shows estimated parameters $\hat{\sigma}_{11}$, $\hat{\sigma}_{12}$ and $\hat{\sigma}_{22}$ for values of $\xi$ ranging from 0 to 1; the
vertical bands represent 95\% confidence intervals. For values of $\xi$ larger than 0.3, the interval estimates appear quite stable. This can
be seen by looking at the relative change in parameter estimate and also width of the confidence intervals. We can see that the estimated extremal
correlation, $\hat{\rho}$, is notably affected by
the measurements in a single station (King Island). As the sub-likelihoods involving that particular station receive increasingly low weights, the
estimates  change
substantially. This behavior is consistent with that observed in our simulated data. To
compare fitted models we also considered the composite likelihood information criterion for model selection discussed in
\cite{Padoan10} and defined by $CLIC(\xi) = - 2 \ellhat(\thetahat_\xi) + \text{tr}\{\hat{H}_\xi^{-1}(\thetahat_\xi)
\hat{K}_\xi(\thetahat_\xi)\}$, where $\hat{J}_\xi$ and $\hat{K}_\xi$ are estimates of the
matrices $H_\xi$ and $K_\xi$ defined in Section \ref{sec:3.1}. We found that the $CLIC(\xi)$ decreases monotonically for $\xi$ in $[0,1]$  -- particularly we
have  we have  $CLIC(0) = 156.6$ and $CLIC(0.3) = 155.9$. This suggests that $\xi>0$ should be preferred  to the usual composite likelihood estimator with
uniform weights with $\xi=0$.

\section{Conclusion and final remarks} \label{sec:conclusion}

This work introduces the D-McLE, a new estimator  obtained by maximizing the weighted composite likelihood function subject to a
discrimination constraint, which entails moving away by a  distance $\xi$ from uniform weights. The D-McLE has
 appealing features from both theoretical and practical viewpoints. First, we found that
the data-adaptive weights render the parameter estimates more stable in the presence of incompatible models compared to classic composite
likelihood approaches with fixed weights. This is clearly seen from our  asymptotic derivations and our numerical simulations confirm this behavior  in finite
samples. Second, the estimated weights, which are a by-product of our procedure, can be used to rank the compatibility of lower-dimensional likelihoods and are
a useful diagnostic tool for model selection. For example, if the $j$th data sub-set  receives an unusually small weight, it is
likely that the corresponding model, $f_j(y_j|\theta)$, is incompatible. Targeted analyses on the anomalous data sub-sets can
lead to improved model assumptions. Third, our approach leads naturally to the algorithm in Section \ref{computing}, which we found to be quite fast and
and easy to implement.

In recent years, high-dimensional estimation has  become a core area of multivariate analysis. We believe that the D-McLE will be valuable as a
remedy to common shortcomings of the classic McLE with fixed weights and MLE when the sample size, $n$, is relatively small compared to the
complexity of the full model. Specifically, the constrained optimization problem (\ref{lagrangian}) is a type of regularization approach where $\lambda_1
D_{KL}(w, w_{unif})$ can be regarded as complexity penalty which promotes sparsity and produces vectors $w$ with many elements close to zero.
Regularization  approaches have proved useful for high-dimensional model  selection \citep{Fan10}. Similarly, in this context, we believe that the design of
new  sparsity-inducing penalty schemes  for likelihood selection would be an interesting direction for further exploration and is high priority in our research
agenda. Findings would be   particularly valuable to  spatial statistics and statistical genetics, where often the large number of sub-likelihood components
poses serious challenges to applicability of composite likelihood methods.

Up to date, not many papers have explored the large-$m$ behavior
of composite likelihood estimators from a theoretical perspective.  \cite{Cox&Reid04} provide useful explanations on the
asymptotic behavior of the pairwise composite likelihood estimator as $m \rightarrow \infty$ and $n$ is fixed; particularly, they discuss how the
presence of strongly correlated partial scores affects the usual convergence rate of McLE. Additional Monte Carlo experiments for the normal location model
defined in
Section \ref{ex:norm} (not reported here)  show that in finite samples the D-McLE can reduce considerably the mean squared error of the uniformly weighted McLE
-- even under fully compatible models. Such accuracy gains are relatively large  when $m$ increases. Developing theoretical insight on this
phenomenon --  and particularly on the interplay between the type of regularization constraint and the mean squared error of the resulting estimator as $m$
increases  --  would represent another exciting future research direction.

\appendix

\section*{Appendix}

\noindent \textit{Lemma 1. If $D_{KL}(\what(\theta), w_{unif}) = \xi$, $\xi\geq 0$, then
$\nabla_\theta \ellhat(\theta)= \sumjm \what_j(\theta) \uhat_j(\theta)$. Therefore,
$\nabla_\theta
\ellhat(\theta)=0$ implies $\nabla_\theta
\alpha_1(\theta) = 0$ with probability going to 1.}

\vspace{.4cm}

\noindent \textit{Proof of Lemma 1.} Let $\alpha_1^\prime(\theta) = \nabla_\theta
\alpha_1(\theta)$. Differentiating both sides of $D_{KL}(\what(\theta), w_{unif}) =
\xi$ gives
$$
0 =  \alpha_1^\prime(\theta)  \dfrac{\sumjm e^{\alpha_1(\theta)
\ellhat_j(\theta)} \ellhat_j(\theta)}{\sumjm e^{\alpha_1(\theta) \ellhat_j(\theta)}}
+ \alpha_1(\theta) \nabla_\theta \ellhat(\theta)
-   \dfrac{\sumjm e^{\alpha_1(\theta) \ellhat_j(\theta)} \{
\alpha_1^\prime(\theta)  \ellhat_j(\theta)+  \alpha_1(\theta)
u_j(\theta)\}}{\sumjm  e^{\alpha_1(\theta)
\ellhat_j(\theta)}},
$$
where $\ellhat(\theta) = \sumjm \what_j(\theta) \ellhat_j(\theta)$. This implies
$\nabla_\theta \ellhat(\theta) =  \sumjm \what_j(\theta) \uhat_j(\theta)$. A calculation
also shows
\begin{equation} \label{eq3}
\nabla_\theta \ellhat(\theta) =  \hat{\alpha}_1^\prime(\theta) \sumjm \what_j(\theta)
\{\ellhat_j(\theta) - \ellhat(\theta)\}^2 +  \sumjm \what_j(\theta)
\uhat_j(\theta),
\end{equation}
Since the first
sum in (\ref{eq3}) is strictly positive with probability one as $n\rightarrow \infty$ and
the second sum
equals zero by the Kullback-Leibler divergence constraint, we have that $\nabla_\theta
\alpha_1(\theta) = 0$ with probability one as $n\rightarrow \infty$.

\subsection*{Proof of Proposition 1}

\noindent The main goal is to show uniform convergence for the composite likelihood
function  $\ellhat(\theta)$. In particular,
\begin{align}
\sup_{\theta \in \Theta} \vert \ellhat(\theta) - \ell(\theta) \vert
&  \leq
\sup_{\theta \in \Theta} \sumjm | \what_j(\theta)\ellhat_j(\theta) -
w_j(\theta)\ell_j(\theta)  | \\
& \leq
 \sumjm \sup_{\theta \in \Theta}  \vert \ellhat_j(\theta) -
\ell_j(\theta) \vert + \sumjm \sup_{\theta \in \Theta}  |\ell_j(\theta)| \ \vert
\what_j(\theta) - w_j(\theta)  \label{eq1}
\vert
\end{align}
The first term in (\ref{eq1}) converges to zero in probability by Condition C2. By the
continuous mapping theorem, also the second term converges to zero.  Next, note that
$\ellhat(\thetahat_\xi)
\geq  \ellhat(\theta^\ast_\xi) = \ell(\theta^\ast_\xi) - o_p(1)$, where the last equality
follows from the weak law of large numbers,
since the latter implies $\ellhat_j(\theta^\ast_\xi) \arrowp \ell_j(\theta^\ast_\xi)$
($1\leq j\leq m$), and the continuous mapping theorem. Hence

\begin{equation}\label{eq2}
\ell(\theta^\ast_\xi) - \ell(\thetahat_\xi) \leq \ellhat(\thetahat_\xi) -
\ell(\thetahat_\xi)+ o_p(1) \leq \sup_{\theta \in \Theta} |\ellhat(\theta) - \ell(\theta)|
+ o_p(1) \rightarrow 0,
\end{equation}
by Condition C2. Since the optimal parameter $\theta^\ast_\xi$ value is unique,
(\ref{eq2}) implies $\thetahat_\xi \arrowp \theta^\ast_\xi$.

\subsection*{Regularity conditions and proof of Proposition 2}

\noindent Let $\nabla$ denote the differential operator with respect to the parameter
vector $\theta \in \Theta \subseteq \mathbb{R}^p$, $\uhat(\theta)=\sumjm \what_j(\theta)
\uhat_j(\theta)$ denotes
the weighted score $p$-vector, with partial scores $\uhat_j(\theta)= n^{-1}\sumin \nabla
\log f_j(y^{(i)}_j|\theta)$, and $H_\xi$,  $K_\xi$ are $p \times p$ matrices defined in
the asymptotic variance
expression (12). Assume (C1)--(C3) given in Proposition 1 and the additional
regularity conditions:
\begin{itemize}
 \item[](C4) the sub-model $f_j(y_j|\theta)$ is three times differentiable in $\theta$, $1 \leq j \leq m$;
 \item[](C5) $\max_{1 \leq k \leq m} E_G| \uhat_k(\theta) |^3$ is upper bounded by a
constant;
 \item[](C6) the smallest eigenvalue of $H_\xi$ is bounded away from zero;
 \item[](C7) the elements of the matrix $K_\xi$ are upper bounded by a constant;
 \item[](C8) the expectation of second-order partial derivatives of $\uhat_k(\theta)$ with
respect to $G$ are upper bounded by a constant for all $\theta$ in a neighborhood of
$\thetaxi$.
\end{itemize}
\noindent By Taylor's Theorem,
there exists a random point $\tilde{\theta}$ between $\thetaxi$ and $\thetahat_\xi$
such that
\begin{align} \label{Taylor}
0 = \uhat(\thetahat_\xi) = \uhat(\thetaxi) + \nabla \uhat(\thetaxi) (\thetahat_\xi  -
\thetaxi )
+ \dfrac{1}{2}(\thetahat_\xi - \thetaxi )^T \nabla^2 \uhat(\tilde{\theta}) (\thetahat_\xi
- \thetaxi ).
\end{align}
For the first term  $\uhat(\thetaxi) = \sumjm \what_j(\thetaxi) \uhat_j(\thetaxi)$ in the
above expansion, the central limit theorem implies that
 $\sqrt{n} \ \uhat_j(\thetaxi)$ converges weakly to a $p$-variate normal distribution with
mean $\mu^\ast_j = E_{G} \uhat_j(\thetaxi)$ and $p\times p$ covariance matrix
$V_j^\ast = -{H_j}^{-1}(\thetaxi)$, for all
$j=1,\dots, m$, where ${H_j}(\theta) = E_G \nabla \uhat_j(\theta)$.  Since
$\ellhat_j(\thetaxi)
\arrowp \ell_j(\thetaxi)$ ($j=1,\dots, m$),
 the continuous mapping theorem implies that $\what_j(\thetaxi)$ converges in probability
to
constants $w_j^\ast = w_j(\theta^\ast_\xi)$ ($j=1,\dots, m$).
Therefore, by Slutsky's theorem we have convergence in distribution of $\sqrt{n}  \
\uhat(\thetaxi)$ to the normal mixture
 $$
\sqrt{n}  \ \uhat(\thetaxi) \arrowd \sumjm w_j(\thetaxi) N_p\{ \mu^\ast_j,
V_j^\ast\}.
$$
such that $\sumjm w_j(\thetaxi) \mu^\ast_j = 0$. For $\nabla \uhat(\thetaxi)$ in the
second term of expansion (\ref{Taylor}), Lemma 1 gives
\begin{align*}
\nabla \uhat(\thetaxi) &= \sumjm  \what_j(\thetaxi) \left[ \nabla \uhat_j(\thetaxi)  +
\hat{\alpha}_1(\thetaxi) \uhat_j(\thetaxi) \uhat_j(\thetaxi)^T
+ \{ \nabla \hat{\alpha}_1(\thetaxi) \} \uhat_j(\thetaxi)^T \ellhat_j(\thetaxi)
\right]
\\
& \arrowp   \sumjm w^\ast_j \left\{ H_j(\thetaxi) + \alpha_1^\ast \mu^\ast_j
{\mu_j^\ast}^T\right\},
\end{align*}
where $\hat{\alpha}_1(\theta)$ is the solution of equation (6) and
$\alpha^\ast_1=\alpha_1(\thetaxi)$ denotes the solution of equation (6) with $\ellhat_j$
replaced by $\ell_j$ and $\theta =
\thetaxi$. Convergence in
probability follows from the continuous mapping theorem since $\ellhat_j(\thetaxi) \arrowp
\ell^\ast_j(\thetaxi)$,
$\uhat_j(\thetaxi) \arrowp \mu^\ast_j$, $\nabla \uhat_j(\thetaxi) \arrowp
H_j(\thetaxi)$. Finally, for the third term of the expansion (\ref{Taylor}) by assumption,
there is
a neighborhood $B$ of $\thetaxi$ and a constant $\kappa$ for which each entry of the array
$E_G \nabla^2  \uhat_k(\theta)<\kappa$ for all $\theta \in B$ and all $k= 1,\dots, p$.
Therefore, $\Vert \nabla^2 \uhat_k(\tilde{\theta}) \Vert$ is bounded in probability by the
law of large numbers. By Proposition 1,  $\thetahat_\xi \arrowp \thetaxi$ and the third
term in the expansion (\ref{Taylor})  is of higher order than the second term, so the
normality result follows by applying Slutsky's Lemma.

\bibliographystyle{abbrvnat}
\bibliography{biblio}

\end{document}